# Local entanglement of electrons in 1D hydrogen molecule

**Ivan P. Christov**
Physics Department, Sofia University, 1164 Sofia, Bulgaria

**Abstract:** The quantum entanglement entropy of the electrons in one-dimensional hydrogen molecule is quantified locally using an appropriate partitioning of the two-dimensional configuration space. Both the global and the local entanglement entropy exhibit a monotonic increase when increasing the inter-nuclear distance, while the local entropy remains peaked at the middle between the nuclei with its width decreasing. Our findings show that at the inter-nuclear distance where stable hydrogen molecule is formed, the quantum entropy shows no peculiarity thus indicating that the entropy and the energy measures display different sensitivity with respect to the interaction between the two identical electrons involved. One possible explanation is that the calculation of the quantum entropy does not account explicitly for the distance between the nuclei, which contrasts to the total energy calculation where the energy minimum depends decisively on that distance. The numerically exact and the time-dependent quantum Monte Carlo calculations show close results.

## 1. Introduction

The entanglement in complex quantum systems manifests itself as non-separability of the many-body quantum state that occurs due to either classical interaction (e.g. Coulomb repulsion between the particles) or as a result of intrinsically quantum non-local effects (e.g. the symmetry of the wave function) [1]. Since in the former the strength of the Coulomb interaction between the particles varies in space it makes sense to ask the question – what the spatial dependence of the entanglement is as a function of the localization of the particles. Basically, the entanglement is quantified using "global" quantum entropy which ultimately boils down to a number where the larger the number the larger the entanglement. It would be expected however that the spatial variations of the interaction strength between the electrons would result in spatially varying entanglement. The intrinsically quantum nonlocality, on the other hand, dictates that the localization of the quantum entropy is to be considered in configuration space where the many-body wave function resides, rather than in physical space. Previous approaches to localize entropy includes the application of Bader's concept of "atoms in molecule" where the local information entropy is calculated in terms of electron density distribution [2]. Other research investigates the localization of the entanglement of two interacting distinguishable particles in configuration space by sampling preliminary calculated reduced density matrix for known wave functions over a uniform grid and next calculating the von Neumann entropy [3]. It is also worth mentioning the use of Tsallis approach to characterize the entropy properties of a single hydrogen molecule as well as of systems of three and more such molecules [4]. From the chemistry viewpoint, the local entanglement entropy considered as information entropy may find application in analysis of electron density and as correlation measure in atoms and molecules where mostly global measures have been used so far [5,6].

The primary goal of this work is to introduce a method for describing the quantum entanglement *locally* by using a localized quantum entanglement entropy as a function of the distance between two one-dimensional hydrogen atoms during their transition from separate species to the formation of a hydrogen molecule at a ground state, and beyond. Unlike in a helium atom where the s-state (opposite spin) electrons occupy essentially the same space, in the hydrogen molecule there is an axis of symmetry and the question about the entanglement fluctuations in space makes more sense, especially for identical particles. Using direct numerical solution of the two-body time-dependent Schrödinger equation it is shown that the reduced density matrix (RDM) for each of the electrons can be found using a set of Monte Carlo walkers which sample the probability distribution in 2D configuration space. The RDM is next used to find the quantum entanglement entropy as an approximation to the von Neumann quantum entropy, named linear entropy, calculated at appropriate regions in configuration space oriented along the separation between the atoms. It is found that the quantum entanglement entropy exhibits maximum in the middle between the atoms, even when those are well separated.

Since, in general, the numerical solution of the Schrödinger equation suffers an exponential scaling of the computational resources with the number of particles, here we also apply the recent Time dependent quantum Monte-Carlo (TDQMC) method which reduces the many-body quantum problem of to a set of problems for particles and waves which are defined in physical space-time [7-10]. This is done in TDQMC by introducing for each electron an ensemble of point-like walkers in space and a concurrent ensemble of guide waves as walkers in the "waves space", where these two ensembles are mutually connected through the set of Schrödinger-type equations, together with guiding equations or with a combined drift-diffusion process. Through the particle-wave dichotomy the particles react to the wave's evolution while, at the same time, the waves experience a back reaction from the particles motion, in a self-consistent manner. The main advantage of using the TDQMC formalism is that the quantum correlations which are due to interaction potentials between the particles can be accounted for in a tractable way, even for many-body systems. The TDQMC method scales almost linearly with the number of particles for bosonic states and scales, in general, no worse than time dependent Hartree-Fock.

The results of the present work would allow one to consider in more detail the quantum information processes of formation of molecules and it can easily be extended to more complex structures such as clusters, nanostructures, etc., in higher spatial dimensions.

## 2. Methods

First, we calculate the ground state of two one-dimensional hydrogen atoms with coordinates of their nuclei $X_1, X_2$, by solving the time dependent Schrödinger equation in imaginary time $t = -i\tau$ (in atomic units):

$$-\frac{\partial}{\partial \tau}\Psi(x_1, x_2, \tau) = H\left(x_1, x_2\right)\Psi(x_1, x_2, \tau) \ , \tag{1}$$

where the Hamiltonian reads:

$$H\left(x_1, x_2\right) = -\frac{1}{2}\left(\frac{\partial^2}{\partial x_1^2} + \frac{\partial^2}{\partial x_2^2}\right) + \sum_{i,j}^{2} V_{e-n}(x_i - X_j) + V_{e-e}(x_1 - x_2) \, , \tag{2}$$

and the imaginary time propagation in Equation 1 ensures that for an arbitrary initial wave function the higher order states are projected out as steady state is established (see e.g. [11]).

To avoid numerical complications due to the singularity of the Coulomb potential at the position of the nuclei, it is assumed that the electron-nuclear and electron-electron interactions are approximated by modified potentials [12]:

$$V_{e-n}(x_i - X_j) = -\frac{1}{\sqrt{1 + (x_i - X_j)^2}} \ ; \tag{3}$$

$$V_{e-e}(x_i - x_j) = \frac{1}{\sqrt{1+\left|x_i - x_j\right|^2}}, \quad (4)$$

where *i,j=1,2*, and where in Equations 1,2 we treat the nuclei as classical particles. This is justified within our approach of gradually changing the distance between the nuclei and next calculating the ground state of the electronic system, without accounting for the nuclear dynamics explicitly. The ground state wave function was found either by solving directly Equation 1 using e.g. the split-step Fourier method or by applying the time-dependent quantum Monte Carlo method which introduces concurrent ensembles of walkers and guide waves defined in physical space where the many-body problem is reduced to a set of coupled Schrödinger-type equations for the guide waves $\varphi_i^k(x_i,\tau)$ for the *i*-th electron; *k=1,2,..., M* denotes the different walkers [7-10]:

$$-\frac{\partial}{\partial\tau}\varphi_i^k(x_i,\tau) = \left[-\frac{1}{2}\frac{\partial^2}{\partial x_i^2} + V_{e-n}\left(x_i, X_j\right) + V_{eff}^k\left(x_i,t\right)\right]\varphi_i^k(x_i,\tau); \quad i=1,2 \quad (5)$$

where $V_{eff}^k\left(x_i,t\right)$ is the effective electron-electron interaction potential given by a Monte Carlo convolution of the true interaction potential $V_{e-e}\left(x_i - X_j\right)$ of Equation 4 and a kernel function $K\left[x_j, x_j^k(t), \sigma_j\right]$, which accounts for the spatial nonlocality experienced by each walker due to the quantum uncertainty:

$$V_{eff}^k\left(x_i,t\right) = \sum_{j\neq i}^{2}\frac{1}{Z_j^k}\sum_{l}^{M} V_{ee}\left[x_i, x_j^l(t)\right] K\left[x_j^l(t), x_j^k(t), \sigma_j\right]; \quad (6)$$

$$K\left[x_j, x_j^k(t), \sigma_j\right] = \exp\left(-\frac{\left|x_j - x_j^k(t)\right|^2}{2\sigma_j^{\,2}}\right), \quad (7)$$

where *i, j=1,2,* and $\sigma_j$ is the characteristic length of spatial nonlocality which is numerically close to the standard deviation of the *j*-th walkers distribution [13] and which variationally minimizes the energy between the limiting cases of pairwise interaction between the walkers ($\sigma_j \to 0$, described by a set of linear equations [8]), and the Hartree-Fock approximation ($\sigma_j \to \infty$) which is essentially nonlinear with respect to the wave functions. Notice that the waves $\varphi_i^k(x_i,\tau)$ considered as random variables in TDQMC do not have their own ontological meaning and similarly the ground-state energies of Equations 5 are not directly related to the ground-state energy of the electron. Notice also that although employing particles and waves, the TDQMC method differs from Bohmian mechanics which is an exact theory relying on the many body wavefunction, and as such it experiences the exponential scaling with the number of physical particles, while the TDQMC method scales almost linearly for opposite-spin electrons.

For imaginary-time propagation the trajectories $x_i^k(\tau)$ are determined by a drift-diffusion process [9]:

$$dx_i^k\left(\tau\right) = v_i^{Dk}d\tau + \eta_i\left(\tau\right)\sqrt{d\tau}, \quad (8)$$

where:

$$v_i^{Dk}\left(\tau\right) = \left.\frac{\nabla_i\varphi_i^k(x_i,\tau)}{\varphi_i^k(x_i,\tau)}\right|_{x_i = x_i^k(\tau)} \quad (9)$$

is the drift velocity whenever birth-death of walkers is applied, and $\eta_i\left(\tau\right)$ is a Markovian stochastic process. At the same time, the walkers $x_i^k(\tau)$ sample the moduli square of the corresponding guide waves $\left|\varphi_i^k(x_i,\tau)\right|^2$.

Here we apply Equations 1-9 to find both the global and the local quantum entanglement entropy of the electrons in 1D hydrogen molecule, where the ground state is entangled due to the Coulomb interaction between the electrons considered as identical particles. As a measure of entanglement, we first use the *global* linear quantum entropy $S(\tau)$ calculated using the reduced density matrix for each of the two electrons in the molecule:

$$S(\tau) = Tr\left[\rho - \rho^2\right] = 1 - \int \rho^2(x, x, \tau)\, dx, \tag{10}$$

where for the exact reduced density matrix we have (for normalized wavefunctions):

$$\rho_i^E(x_i, x_i', \tau) = \int \Psi(x_i, x_2, \tau)\Psi^*(x_i', x_2, \tau) dx_2 \tag{11}$$

On the other hand, the guide waves provided by the TDQMC method for the *i*-th electron (Equation 5) can be used to efficiently calculate the one-body density matrix considered as covariance matrix for the random variables $\varphi_i^k$ [14], without the need to calculate the density matrix of the whole system:

$$\rho_i^{TDQMC} = E\left(\varphi_i^*\varphi_i\right) = \int P\left[\varphi_i\right]\varphi_i^*\varphi_i D\varphi_i^* D\varphi_i \tag{12}$$

where the random state $\varphi_i^k$ is defined in terms of the probability distribution $P\left[\varphi_i\right]$. Assuming that, according to the particle-wave dichotomy [9], the probability distribution of the waves $\varphi_i^k$ corresponds to the distribution of the walkers $x_i^k$, we arrive at the TDQMC reduced density matrix:

$$\rho_i^{TDQMC}(x, x') = \frac{1}{M}\sum_{k=1}^{M}\varphi_i^{k*}(x)\varphi_i^k(x'), \tag{13}$$

which, in fact, is nothing more than a Monte-Carlo sum representation of the integral in Equation 12. It is seen from Equation 13 that the density matrix is normalized to unity trace as long as the states $\varphi_i^k$ are normalized.

## 3. Results

Henceforth we shall assume that the ground state has been established and we will therefore omit the time variable $\tau$. It is seen from Equation 12 that the RDM provided by the TDQMC method is calculated through the wavefunctions of the different walkers (the upper index *k*). To compare the TDQMC RDM with the exact results for 1D hydrogen molecule we should calculate the exact density matrix in a similar way. If we assume that the two-body probability density $|\Psi(x_1, x_2)|^2$ is sampled by another set of Monte Carlo walkers ($x_1'^k, x_2'^k$), we may easily calculate the reduced wave functions:

$$\psi_1^k\left(x_1'\right) = \mathrm{N}_1\Psi\left(x_1', x_2'^k\right), \tag{14}$$

$$\psi_2^k\left(x_2'\right) = \mathrm{N}_2\Psi\left(x_1'^k, x_2'\right), \tag{15}$$

where $\mathrm{N}_{1,2}$ are normalization factors. The wave functions in Equations 14, 15 can be next used to calculate the reduced density matrix (also named the conditional density matrix [15]) which can also be considered exact since it is based on the numerically exact two-body wave functions in Equations 14,15:

$$\rho_i^C\left(x_i, x_i'\right) = \frac{1}{M}\sum_{k=1}^{M}\psi_i^{k*}(x_i)\psi_i^k(x_i') \tag{16}$$

Although the calculation of the reduced density matrix in Equations 14-16 seems awkward, the use of walkers is important in calculating the local entanglement below.

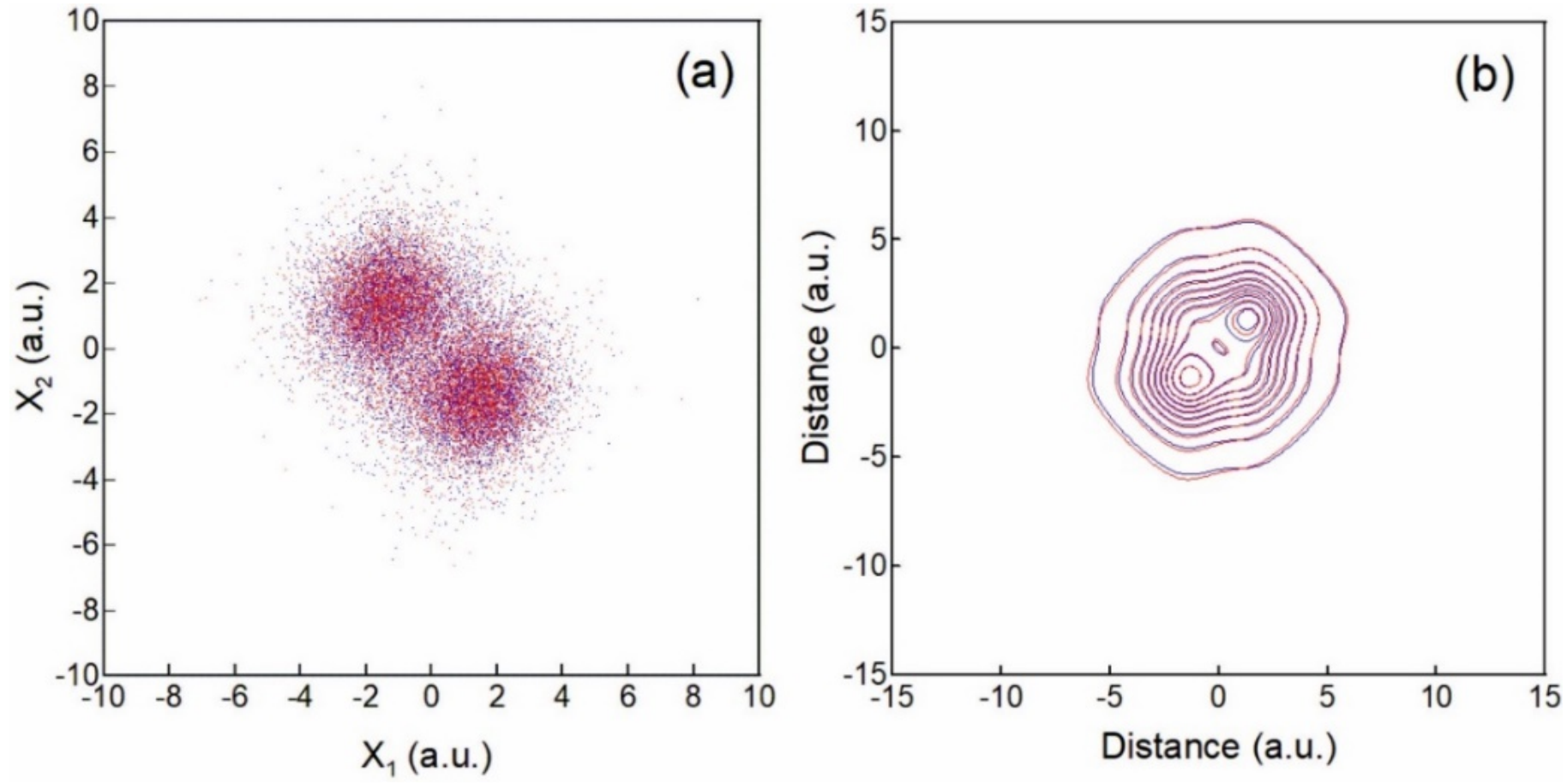

**Figure 1.** Walker distribution in configuration space for molecule of two 1D hydrogen atoms at a distance 3a.u. – **(a)**; contour maps of the reduced density matrix – **(b).** Blue lines and points – exact result, red lines and points – from TDQMC.

We start with the calculation of the walker's distributions and the reduced density matrices for a fixed inter-nuclear distance $\left|X_1 - X_2\right| = 3a.u.$ Figure 1b presents the contour-line images of the exact RDM $\rho^E\left(x,x'\right)$ from Equation 11 (blue line) as compared to the TDQMC RDM $\rho^{TDQMC}\left(x,x'\right)$ from Equation 13 (red line) where the two electron distributions are at their ground state (Figure 1a). It is seen that there is a good correspondence between the exact and the TDQMC results.

Next, we calculate the ground-state energy of the two-atom configuration as function of the distance between the nuclei [16]:

$$E_1 = \frac{1}{M}\sum_{k=1}^{M}\left[\sum_{i=1}^{2}\left[-\frac{1}{2}\frac{\nabla_i^2\varphi_i^k(x_i^k)}{\varphi_i^k(x_i^k)} + V_{e-n}(x_i^k)\right] + \sum_{i>j}^{2} V_{e-e}(x_i^k, x_j^k)\right]_{\substack{x_i^k = x_i^k(\tau) \\ x_j^k = x_j^k(\tau)}} + V_{n-n}\left(X_1 - X_2\right) \qquad (17)$$

for the TDQMC calculation, and:

$$E_2 = \iint\left[-\frac{1}{2}\Psi^*\left(x_1,x_2\right)\left(\frac{\partial^2}{\partial x_1^2} + \frac{\partial^2}{\partial x_2^2}\right)\Psi\left(x_1,x_2\right)\right]dx_1dx_2$$
$$+\iint\left[\sum_{i,j}^{2} V_{e-n}(x_i - X_j) + V_{e-e}(x_1 - x_2)\right]\left|\Psi\left(x_1,x_2\right)\right|^2 dx_1dx_2$$
$$+V_{n-n}\left(X_1 - X_2\right) \qquad (18)$$

for the exact two-electron problem. In Equations 17,18 $V_{n-n}\left(X_1 - X_2\right)$ is the soft nucleus-nucleus potential:

$$V_{n-n}\left(X_1 - X_2\right) = \frac{1}{\sqrt{0.5 + \left(X_1 - X_2\right)^2}} \qquad (19)$$

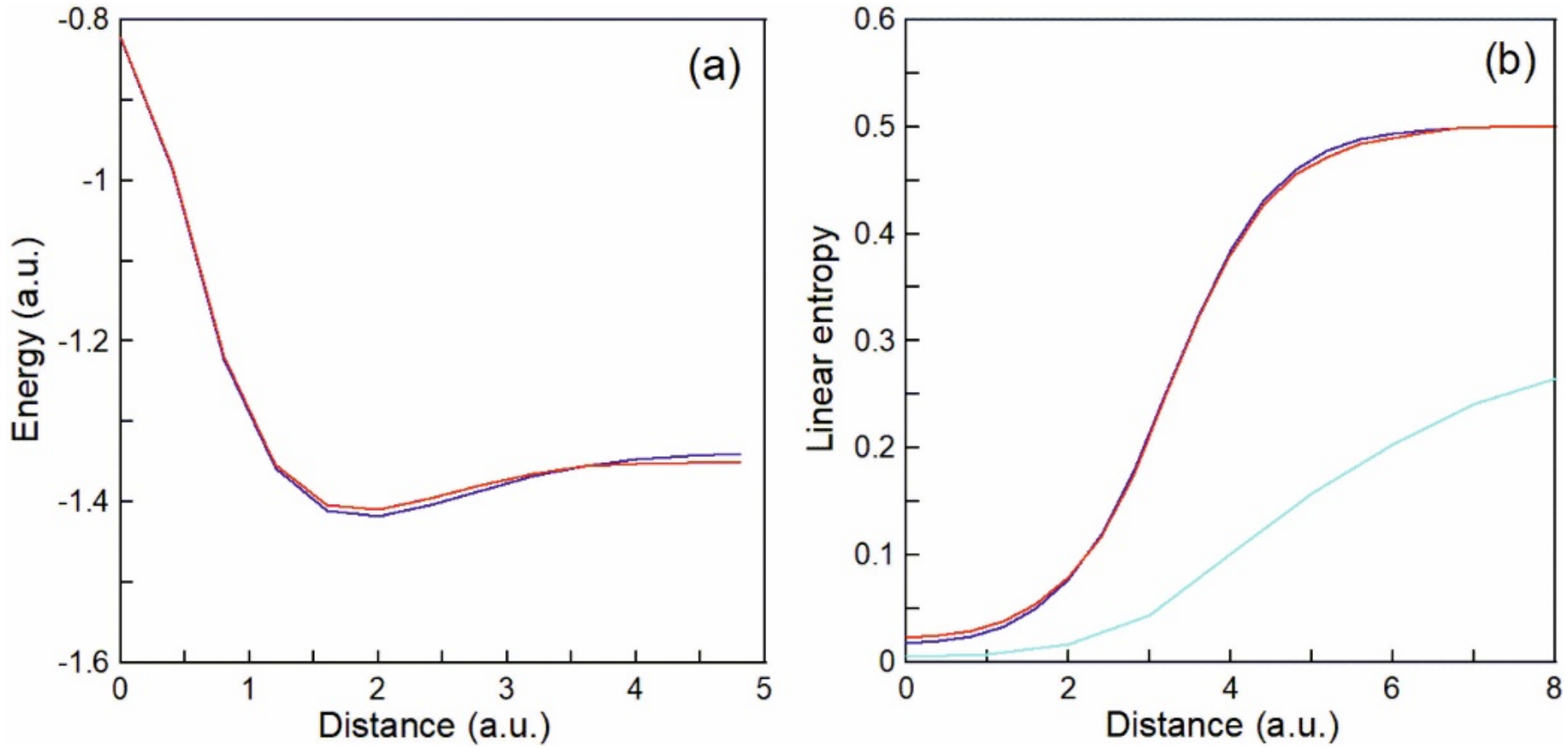

**Figure 2.** Ground state energy of hydrogen molecule – **(a)** and global linear entropy – **(b)** as function of the internuclear distance. Blue lines – exact result, red lines – from TDQMC. The green line in **(b)** shows the peak value of the local entropy, multiplied by 10.

Figure 2a shows the minimum of the energy which occurs at a distance $|X_1 - X_2| \sim 2a.u.$ between the nuclei where a stable hydrogen molecule is formed for both the exact (blue line) and for the TDQMC (red line) calculations. Figure 2b shows with red and blue lines the global linear entropy as predicted by eq.10 as a function of the inter-nuclear distance, for the two electrons considered as identical particles (see figure 1(a)). It is seen that the TDQMC prediction from Equations 10,13 (blue line) is in good agreement with the numerically exact result of Equations 10,11 (blue line). The red and the blue curves in Figure 2b are qualitatively similar to the result in [17] where the configuration interaction method has been used to estimate the global entanglement of the hydrogen molecule as function of the inter-nuclear separation.

Next, we focus on our approach to calculate the *local* entanglement at different regions in configuration space $(x_1, x_2)$. Figure 3a depicts the partition we use here represented as a set of square regions along the diagonal of the walker's distribution of Figure 1a where the global reduced density matrix of Equation 13 can be represented as a sum:

$$\rho_i^{\mathrm{TDQMC}}(x,x') = \sum_{m=0}^{N} \frac{M_m}{M} \rho_i^m = \frac{1}{M} \sum_{k=1}^{M} \varphi_i^{k*}(x) \varphi_i^k(x'), \qquad (20)$$

where $N$ is the total number of square regions in Figure 3a, and $M_m$ is the number of walkers within the $m$-th square region. The $M_0 = M - \sum_{m=1}^{N} M_m$ walkers outside all square regions form one additional region, $m = 0$. With this region every walker belongs to exactly one region, so that the weighted sum reproduces the reduced density matrix, and the additional region is not used in the local entropies below. Notice that although the density matrix can always be represented as the sum in Equation 20, the quantum entropy, which is a nonlinear function of the density matrix, is not additive. In order to cover the density distribution adequately we use 50 regions in Figure 3a with a total of 200 000 walkers for each electron. Then we calculate the local entanglement using *local* linear entropy introduced according to:

$$S_i^m = \int \left[ \rho_i^m(x,x) - \rho_i^{m2}(x,x) \right] dx \qquad (21)$$

where the standard normalization for the local density matrices $Tr\left[\rho_i^m\right] = 1$ is in place. The local entanglement in Equation 21 does not depend on the number of walkers in the total sample (Figure 1a) once this number is large. It depends on the size of the square regions, since a square region restricts the positions of both electrons, and we keep this size fixed when the distance between the nuclei is varied. Clearly the choice of the partition regions shown in Figure 3a is not unique, however, it proves to be a good choice in our case where the distance between the nuclei is to be varied.

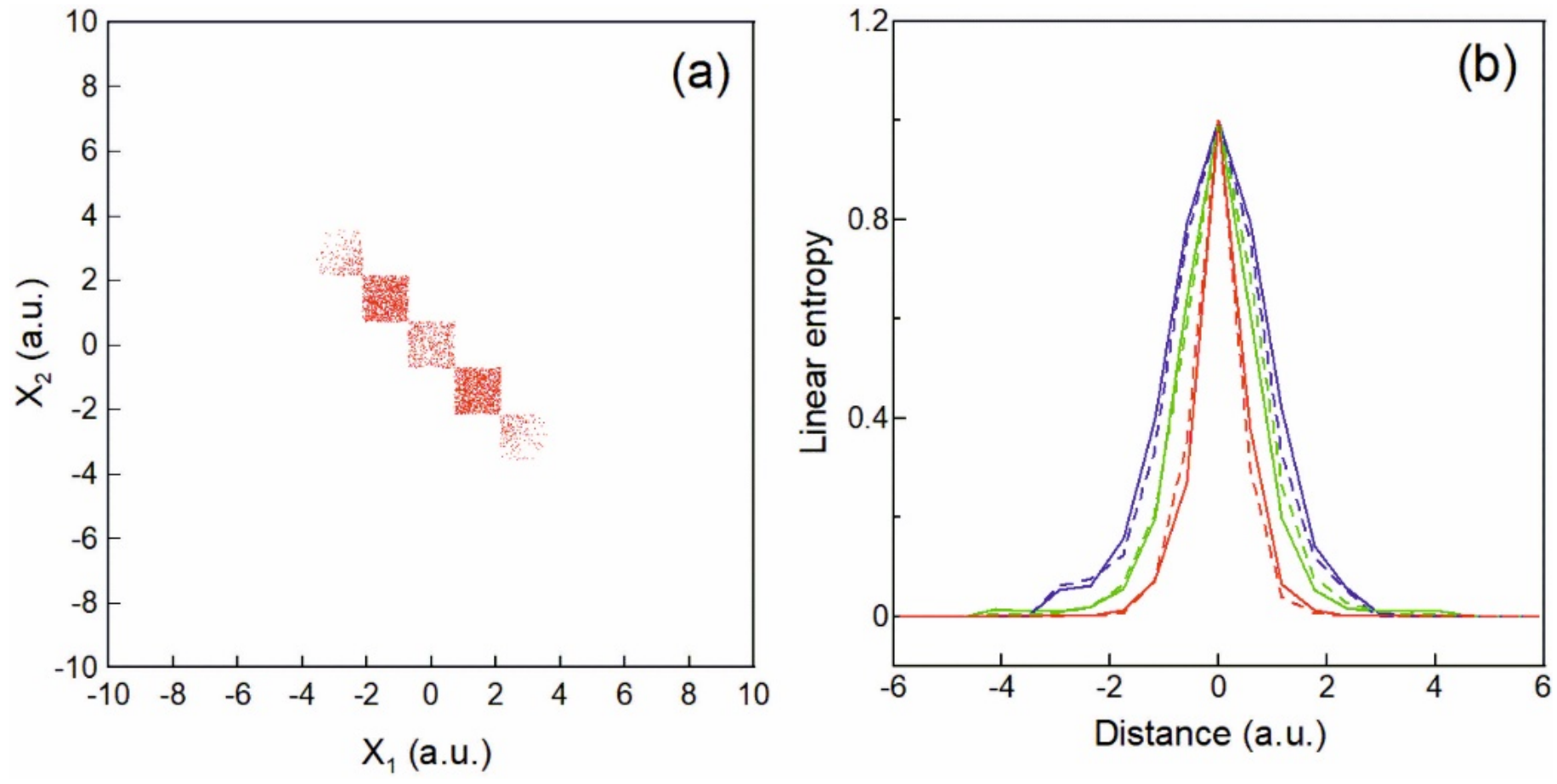

**Figure 3.** Partition of the 2D configuration space used in the calculations: **(a)** - square regions, the rest of the configuration space forms the additional region *m=0* of Eq.(20); **(b)** - local linear entropy for different inter-nuclear distance d: d=0 - blue line, d=3a.u. - green line, d=5a.u. - red line. Solid lines - exact results, dashed lines TDQMC results.

Figure 3b shows the local linear entropy (from Equation 21), normalized to unity, for different inter-nuclear distance $d = |X_1 - X_2|$, from the TDQMC calculation (dashed lines) to be compared with the exact calculation (solid lines) where the representation in Equation 20 is applied also to the conditional density matrix (Equation 16). It is seen from Figure 3b that the maximum of the local entropy is located right in the middle between the two hydrogen atoms while its width decreases by almost a factor of two between distances between the atoms d=0 (blue line) and d=5a.u. (red line). Also, it was found that the peak value of the local entropy (without normalization) monotonically increases from d=0 to d=5a.u. as is shown with a green line in Figure 2b. These findings indicate that within our approach the local entanglement of the two electrons is localized in the middle between the two nuclei, it becomes narrower with increasing the nucleus-nucleus separation, while its peak value increases.

## 4. Conclusions

Here, the global and the local entanglement of the electrons in a simple hydrogen molecule is quantified as a function of the distance between their nuclei. The results from the exact numerical solution of the two body Schrödinger equation are compared with those from the time-dependent quantum Monte Carlo method which essentially reduces the quantum many-body problem from configuration space to physical space. The reduced density matrix of each of the two identical electrons is calculated in the standard way, which shows a monotonic increase of the global linear quantum entropy as the two hydrogen atoms move apart. To address the question of what the local dependence of the quantum entropy would be, we design a special partition of part of the configuration space along the orientation direction of the two-electron configuration and employ a set of particles (walkers) to sample the two-body probability distribution at the different regions. As a result, the "global" density matrix can be represented as a sum of "local" density matrices attached to each of the different regions in configuration space, where we were able to quantify the quantum entropy and hence the entanglement locally. Our findings reveal first that both the exact and the TDQMC calculation provide close results for the reduced density matrix while at the same time the TDQMC method scales much more favorably as compared to the exact solution. The essential result here is that unlike in atoms the entanglement of identical electrons in hydrogen molecule becomes localized not at the positions of the nuclei but in the middle between those, as the two atoms move apart. That result was confirmed by the two independent methods used and it is somewhat unexpected in the light of the spreading of the wave function of each electron in the molecule which makes it less localized in space.

**Funding:** This research is based upon work supported by the Air Force Office of Scientific Research under award number FA8655-22-1-7175, and by the Bulgarian Ministry of Education and Science as a part of National Roadmap for Research Infrastructure, grant number D01-298/17.12.2021 (ELI ERIC BG).